\documentstyle[aps,multicol]{revtex}

\begin{document}

\begin{multicols}{2}

{\bf Comment on ``Solution of Classical Stochastic One-Dimensional
Many-Body Systems''}

In a recent Letter, Bares and Mobilia\cite{bares} 
proposed the method to find solutions of the
stochastic evolution operator $H=H_0 + {\gamma\over L} H_1$ with a non-trivial
quartic term $H_1$.
They claim, 
``Because of the conservation of probability, an analog of the Wick theorem
applies and all multipoint correlation functions can be computed.''
Using the Wick theorem, they 
expressed the density correlation functions as solutions of a closed
set of integro-differential equations.

In this Comment, however, we show that 
applicability of Wick theorem is restricted to the case $\gamma = 0$ only. 
To discuss this point, let us consider the generating function of 
correlation functions 
$
Z[\xi;t] \equiv \langle \tilde \chi| \exp [
\sum_q \xi_q a_q ] e^{-Ht}|\rho \rangle
$ 
introduced by Santos {\it et al.}\cite{sss}
Here $a_q(a_q^\dag)$ is a $q$-mode annihilation (creation) operator 
and 
$\xi_q$ is a Grassmann number with anticommutation relations
$
\{\xi_q , a^\dag_p\} = \{ \xi_q, a_p\} = \{ \xi_q, \xi_p\} = 0
$
for all momentum indices $p,q$. 
$\langle \tilde \chi |$ and $| \rho \rangle$ are the left vacuum and the initial state,
respectively, belonging to the  even sector of Hilbert space\cite{sss}.
For the initial state considered in Ref.\cite{bares},
the generating function $Z$ is
\begin{equation}
Z[\xi;0] = \exp \left [ \sum_{q>0} {{\mu^2 \cot({q \over 2})} \over
{1 + \mu^2 \cot^2({q \over 2})}} \xi_q \xi_{-q} \right ]
\label{initial}
\end{equation}
with $\mu\equiv \rho /(1 -\rho)$ \cite{sss}.
As mentioned in Ref. \cite{sss}, the Wick theorem at $t=0$ follows from the
Gaussian form of the generating function.
The applicability of Wick theorem for later time can be tested by
checking that 
the generating function is a Gaussian for later time or not.

Now let us suppose that $Z$ be a Gaussian with the form 
$
Z[\xi;t_0] = \exp [ \sum_{q>0} f(q,t_0)\xi_q \xi_{-q}  ]
$ at time $t = t_0$.
For simplicity we assume that $f$ is an odd function of $q$
as in Eq.(\ref{initial}).
After an infinitesimal time increment $dt$, the generating function becomes
$
Z[\xi;t_0+dt] 
= Z[\xi;t_0]  
+ dt \langle \tilde \chi |
 [H , e^{\sum_q \xi_q a_q }] e^{-Ht_0} |\rho \rangle,
$ 
where $[,]$ denotes the  usual commutation relation and we use the fact
that $\langle \tilde \chi | H = 0$.
Using the property of the left vacuum ($\langle
\tilde \chi | a_q^\dag = \cot({q\over 2}) \langle \tilde \chi |
a_{-q} $, see Ref.\cite{sss}) 
and the commutation relations
$
[a_q^\dag, \zeta] =
 - \xi_q \zeta,~~
 [a_q,\zeta] = 0,
$
($\zeta\equiv\exp[\sum_q \xi_q a_q ]$), we obtain  
the time evolution of the generating function.
In the absence of the quartic term ({\it i.e.} $\gamma=0$), the generating function 
evolves as
\begin{eqnarray}
&&Z[\xi;t_0 + dt] =
\exp \Biggl \{ \sum_{q>0} \xi_q \xi_{-q} \times \nonumber\\
&\times&\left [
f(q,t_0) - dt ( \nu(q) + \bar \nu(q) ) f(q,t_0) + 2 \epsilon dt \sin q 
\right ] \Biggr \},
\label{z0}
\end{eqnarray}
where  $\nu(q) = \omega(q) + 2 \epsilon \sin q \cot({q \over 2})$\cite{bares}.
This implies that $Z$ remains a Gaussian at later time, if $Z[\xi;0]$ is a 
Gaussian (see Eq.(\ref{initial})).
As a byproduct, we obtain a differential equation (DE) 
for $f$.
Since the DE for $f$ preserves the parity in $q$, $f$ is an odd function for
all $t$.
In the presence of the quartic term ({\it i.e.} $\gamma\neq0$), however, 
the situation should be modified significantly.
After performing the commutation relation between quartic term and $\zeta$,
we find
\end{multicols}
\widetext
\begin{eqnarray}
&&Z[\xi;t_0]^{-1}\langle \tilde\chi|[H_1 , \zeta] 
e^{-Ht_0}| \rho \rangle = 2 \sum_{p,q} (\cos q \cos p - 1) \cot ({p \over 2})
\xi_q \xi_{-q} f(p,t_0)f(q,t_0)
+ 2
 \sum_{p,q}  \sin q \sin p \cot ({q \over 2})
\xi_q \xi_{-q} f(q,t_0) f(p,t_0) \nonumber \\
&&- \left [
\sum_p \sin p f(p,t_0) \right ] \left [ \sum_{q} \sin q \xi_q
\xi_{-q} \right ]
 + \sum_{q_1,q_2,q_3,q_4} 
\cos (q_1 + q_3) \delta(q_1+q_2+q_3+q_4)
f(q_3,t_0) f(q_4,t_0) \xi_{q_1} \xi_{q_2} \xi_{q_3} \xi_{q_4} 
\nonumber\\
&&+\sum_{q_1,q_2,q_3,q_4}
 [\cos(q_2+q_3) - \cos(q_1+q_3)] \delta(q_1+q_2+q_3+q_4) 
\cot({q_1 \over 2}) f(q_1,t_0) f(q_3,t_0)
f(q_4,t_0)\xi_{q_1} \xi_{q_2}
\xi_{q_3} \xi_{q_4}. 
\label{4op}
\end{eqnarray}
\begin{multicols}{2}
\noindent
Although $Z[\xi;t_0]$ is a Gaussian initially, the quartic terms in 
Eq.(\ref{4op}) prevent
it from remaining a Gaussian at later time, thus break the 
Wick theorem.

If we ignore the 
quartic term in the generating function, we indeed find the integro-differential
equation of $g(q,t)$($\equiv f(q,t) \cot ({q \over 2} )$) as given
by Eq.(7) in Ref.\cite{bares}. Therefore,
it may be considered as a Hartree-Fock type approximation\cite{ppk}. 
Since the quartic term is irrelevant in the model considered, 
Eq. (7) in Ref.\cite{bares} predicts the correct (leading)
long time behavior of the density and the correlation
function.
However, for the systems with the relevant non-quadratic 
terms in the evolution operator, the validity of the 
method proposed in Ref.\cite{bares} is doubtful.

This work was supported by the KOSEF through Grant No. 
981-0202-008-2 and by CTP at SNU.

Su-Chan Park${}^1$, Jeong-Man Park${}^2$ and Doochul Kim${}^1$ 

${}^1$ Department of Physics, Seoul National University, Seoul, Korea

${}^2$ Department of Physics, The Catholic University of Korea, Puchon, Korea

\end{multicols}
\end{document}